\documentclass[aps,prl,10pt,reprint,groupedaddress,superscriptaddress]{revtex4-2}
\pdfoutput=1

\usepackage{graphicx}
\usepackage{dcolumn}
\usepackage{bm}
\usepackage[nointegrals]{wasysym}
\usepackage[unicode=true,colorlinks=true]{hyperref}
\hypersetup{linkcolor=blue,citecolor=blue,urlcolor=blue}
\usepackage{graphicx}
\usepackage{epstopdf}
\usepackage[dvipsnames]{xcolor}
\usepackage{amsmath}
\usepackage{amsfonts}
\usepackage{amssymb}
\usepackage{bbm}
\usepackage{wasysym}
\usepackage{csquotes}
\usepackage[normalem]{ulem}

\usepackage{booktabs}
\usepackage{braket}

\usepackage{dsfont}
\usepackage{slashed} 

\usepackage{tikz}

\usepackage{pdfpages} 
\usepackage{pgffor} 

\makeatletter
\AtBeginDocument{\let\LS@rot\@undefined}
\makeatother

\def\supplementfilename{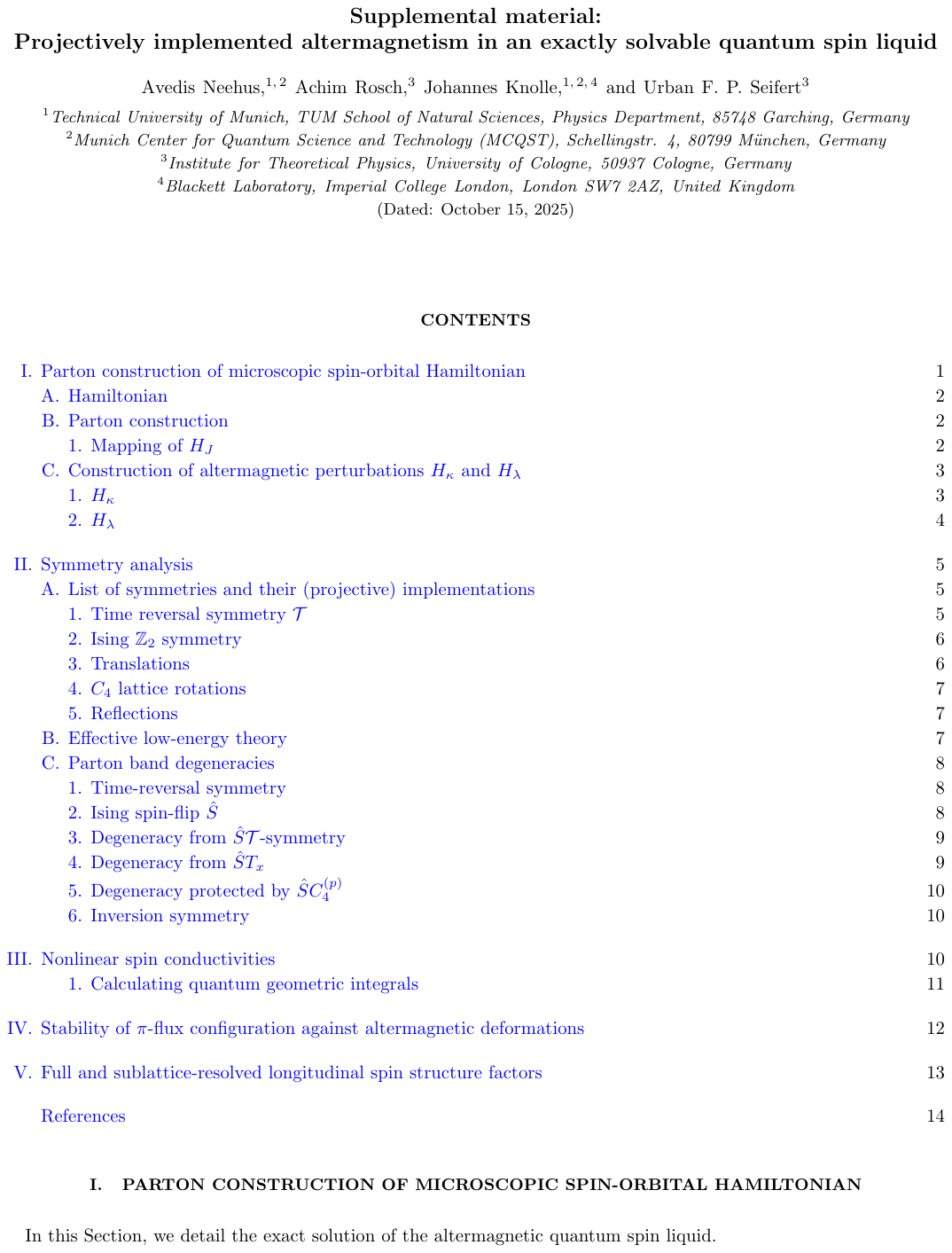}

\pdfximage{\supplementfilename}
\def\numbersupplementpages{\the\pdflastximagepages}

\newif\ifarXiv
\arXivtrue

\definecolor{cadmiumgreen}{rgb}{0.0, 0.40, 0.04}
\definecolor{orange}{RGB}{255,127,0}
\definecolor{blue2}{RGB}{33,114,173}

\renewcommand{\Im}{\operatorname{Im}}

\newcommand{\iu}{\mathrm{i}} 
\newcommand{\eu}{\mathrm{e}} 
\newcommand{\du}{\mathrm{d}} 
\newcommand{\hc}{\mathrm{h.c.}} 

\newcommand{\bvec}[1]{{\bm{#1}}}

\newcommand{\Ztwo}{\mathbb{Z}_2}
\newcommand{\Uone}{\mathrm{U(1)}}

\newcommand{\ph}{\mathrm{\ph}}

\newcommand{\sop}{\mathsf{s}}
\newcommand{\oop}{\mathsf{t}}

\begin{document}
\preprint{APS/123-QED}
\title{Projectively implemented altermagnetism in an exactly solvable quantum spin liquid}

\newcommand{\TUM}{\affiliation{Technical University of Munich, TUM School of Natural Sciences, Physics Department, 85748 Garching, Germany}}
\newcommand{\MCQST}{\affiliation{Munich Center for Quantum Science and Technology (MCQST), Schellingstr. 4, 80799 M{\"u}nchen, Germany}}
\newcommand{\Imperial}{\affiliation{Blackett Laboratory, Imperial College London, London SW7 2AZ, United Kingdom}}
\newcommand{\Cologne}{\affiliation{Institute for Theoretical Physics, University of Cologne, 50937 Cologne, Germany}}

\author{Avedis Neehus} \TUM \MCQST
\author{Achim Rosch} \Cologne
\author{Johannes Knolle} \TUM \MCQST \Imperial
\author{Urban F. P. Seifert} \Cologne

\date{October 15, 2025}

\begin{abstract}
Altermagnets are a new class of symmetry-compensated magnets with large spin splittings.
Here, we show that the notion of altermagnetism extends beyond the realm of Landau-type order: we study exactly solvable $\mathbb{Z}_2$ quantum spin(-orbital) liquids (QSL), which simultaneously support magnetic long-range order as well as fractionalization and $\mathbb{Z}_2$ topological order.
Our symmetry analysis reveals that in this model three distinct types of ``fractionalized altermagnets (AM$^*$)''  may emerge, which can be distinguished by their residual symmetries.
Importantly, the fractionalized excitations of these states carry an emergent $\Ztwo$ gauge charge, which implies that they transform \emph{projectively} under symmetry operations.
Consequently, we show that ``altermagnetic spin splittings'' are now encoded in a momentum-dependent particle-hole asymmetry of the fermionic parton bands.
We discuss consequences for experimental observables such as dynamical spin structure factors and (nonlinear) thermal and spin transport.

\end{abstract}
\maketitle

``Altermagnetism'' has been introduced to characterize a class of magnetic systems  with vanishing net magnetization, where electronic bands exhibit a large spin splitting even in the absence of spin-orbit coupling \cite{ahn19,Hayami2019, Libor2020, Yuan2020, Ma2021, Mazin_2021, libor22b,libor22a, Libor2022c}.
The absence of a net magnetization is protected by a combination of spin rotations with lattice symmetry operations other than translations and inversions.

While altermagnets are intensely pursued in the field of spintronics \cite{Bai24}, there has also been considerable interest into the interplay of correlation physics and altermagnetism \cite{Das_2024,leeb24,sobral24,pupim2024,Banerjee24, steward25, Brekke23,consoli25,eto2025, heung2024, ferrari24, roig24}. 
Here, the altermagnetic phase may be understood in terms of a \mbox{(semi-)}classical Landau theory framework, where both the antiferromagnetic Néel order parameter and some higher-order multipole of the magnetization become finite \cite{Hayami2019,bhowal24,mcclarty24,mcclarty25}.

However, strong quantum fluctuations in quantum spin systems can also stabilize ``quantum spin liquids (QSL)'' which feature a long-ranged entanglement structure, supporting excitations with fractionalized quantum numbers and topological order \cite{savary17}.
As such, they lie outside a Landau theory description of symmetry-broken states and are instead described in terms of deconfined phases of emergent gauge theories. 

In this work, we consider the emergence of altermagnetism in fractionalized systems.
Specifically, we investigate a Kugel-Khomskii-type spin-orbital model \cite{kugelkhomskii_1982,churchill25} which can be mapped onto an exactly solvable $\mathbb{Z}_2$ lattice gauge theory with spinless fermions as fractionalized (parton) degrees of freedom.
Adding perturbations to this model that respect ``altermagnetic'' symmetry requirements gives rise to states that \emph{simultaneously} support magnetic long-range order \emph{and} fractionalization with $\Ztwo$ topological order.
Succinctly, this state can be called a ``fractionalized altermagnet (AM$^\star$)'' in analogy to ``AF$^\star$'' (fractionalized antiferromagnets) \cite{nayak00,senthil04,ghaemi06}.
A similar terminology was recently introduced for quantum-disordered orbital altermagnets \cite{sobral24}, but we emphasize that these retain full spin-rotation invariance.

A key property of our model is the fact that its fermionic excitations (``partons'') carry a non-trivial charge under the emergent $\Ztwo$ gauge group and, therefore, physical symmetry operations act \emph{projectively} on these degrees of freedom. 
We show that, crucially, the projective implementation of symmetries implies that the momentum-dependent time-reversal symmetry breaking is now encoded in a broken particle-hole degeneracy of these fermions.
This should be contrasted with non-fractionalized altermagnets, exhibiting a momentum-dependent splitting of otherwise spin-degenerate bands.
While single-parton spectra are not directly observable, we establish characteristic experimental signatures of fractionalized altermagnets.

\begin{figure}[!htb]
    \centering
    \includegraphics[width=\linewidth]{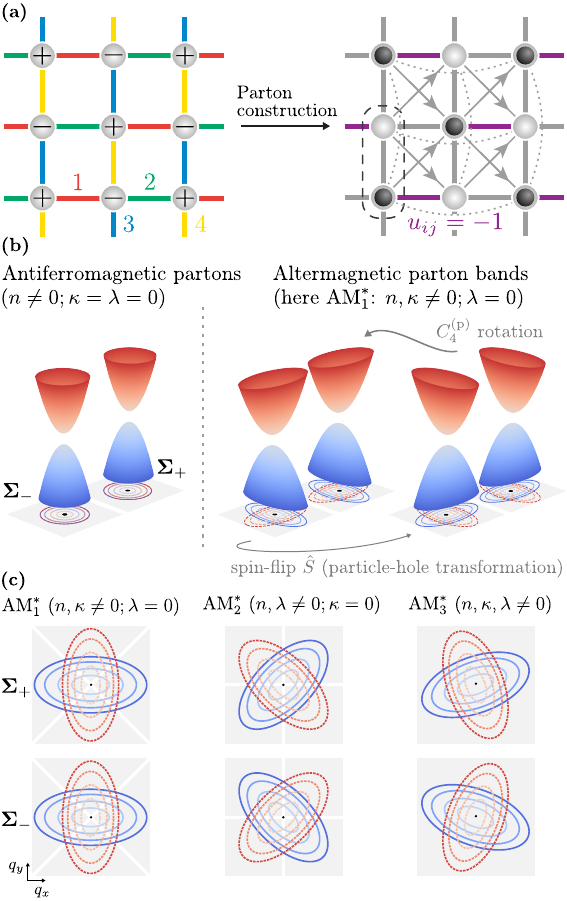}
    \caption{\label{fig:introfic}(a) Square-lattice spin-orbital model with bond-dependent interactions. The staggered out-of-plane magnetization $\langle \sop^z_i \rangle \propto n (-1)^i$ is indicated by ``$+$'' and ``$-$'' symbols. The model is mapped onto an exactly solvable square-lattice $\Ztwo$-gauge theory coupled to spinless fermions, with antiferromagnetism corresponding to charge-density wave order. The purple lines, grey arrows and dotted lines indicate the chosen gauge for the $\pi$-flux ground state of the $\Ztwo$ gauge field, the next-nearest neighbor hopping in $H_\lambda$ and  third-nearest neighbor hopping described by $H_\kappa$. (b) Low-energy parton band structures near the Dirac points $\bvec{\Sigma}_\pm$ for antiferromagnetic and altermagnetic states: in the latter, the bands are symmetric under a combination of spin-flip (particle-hole transformation of the parton bands) and lattice rotations. (c) Illustration of parton bands (in red (blue): constant-energy cuts of particle (hole) band) in the three distinct fractionalized altermagnetic phases for small momenta $\bvec{q}$ near the Dirac points $\bvec{\Sigma}_\pm$. White lines denote nodal lines with residual particle-hole degeneracy.}
\end{figure}

{\em Spin-orbital liquids and projectively implemented symmetries.---} Our starting point is a spin-orbital model on the square lattice with an XY coupling in the spin sector and a Kitaev-type bond-dependent interaction in the orbital sector \cite{yaolee11,chulli20,ufps20,chulli21,vijayvargia23}
\begin{equation}
       H_J = J\sum_{\langle ij \rangle_\gamma} (\mathsf{s}^x_{ i} \mathsf{s}^x_{ j} + \mathsf{s}^y_{ i}  \mathsf{s}^y_{ j}) (\oop^\gamma_i \oop^\gamma_j) \label{eq:HJ}
\end{equation}
where $(\sop^{\alpha})_{\alpha={x,y,z}}$ and $(\oop^\gamma)_{\gamma = 1,2,3,4} = (\oop^x,\oop^y,\oop^z,\mathds{1})$ are Pauli matrices acting in the spin and orbital sector, and $\langle ij \rangle_\gamma$ labels a nearest-neighbor bond of type $\gamma=1,2,3,4$ as illustrated in Fig.~\ref{fig:introfic}(a). {Note that the Hamiltonian $H_J$ belongs to a large class of (generalized) Kugel-Khomskii models for exchange between spin-1/2 and a twofold-degenerate orbital degree of freedom \cite{kugelkhomskii_1982,churchill25}.}

{The Hamiltonian \eqref{eq:HJ} can be solved exactly by introducing 6 Majorana partons $(b^1_i, \dots b^4_i, c^x_i, c^y_i)$ to represent the spin and orbital operators as \begin{align}
        \begin{pmatrix}
           \mathsf{s}^x_i\\
          \mathsf{s}^y_i\\ 
          \mathsf{s}^z_i
         \end{pmatrix} = -\frac{\iu}{2}   
    \begin{pmatrix}
           c^x_i\\
          c^y_i\\ 
          b^4_i
         \end{pmatrix}
         \times
             \begin{pmatrix}
           c^x_i\\
          c^y_i \\
          b^4_i
         \end{pmatrix} \quad \text{and} \quad 
         \oop^\alpha_i  = -\frac{\iu}{2} \epsilon^{\alpha\beta\gamma}b^\beta_i b^\gamma_i. \label{eq:sop-top-majorana}
\end{align} A local fermion parity constraint $D_i = -\iu b^1_ib^2_ib^3_ib^4_ic^x_ic^y_i = -1$ is imposed to project out redundant degrees of freedom. Inserting into \eqref{eq:HJ}, one observes that $u_{ij} = \iu b^\mu_i b^\mu_j$ on a nearest-neighbor link $\langle ij \rangle$ of type $\mu=1,2,3,4$ are conserved quantities which anticommute with $D_i$. Hence, the $u_{ij}$ form a $\Ztwo$ gauge field, with gauge transformations generated by the fermion parity constraint (see also \cite{SuppMat,ufps20} for details). It is further convenient to combine the $c^x,c^y$ Majorana fermions into complex fermions $f_i = (c^x_i + \iu c^y_i)/2$ ($f_j = (\iu c^x_j - c^y_j)/2$) on A (B) sublattices. $H_J$ then becomes a theory of complex spinless fermions coupled to a static $\Ztwo$ gauge field,} 
\begin{equation} \label{eq:HJg}
    \tilde{H}_J = 2 J \sum_{\langle ij \rangle} u_{ij} \left( f_i^\dagger f_j + \hc \right).
\end{equation}
Note that in this basis, {$\sop^z_i = 1- 2f_i^\dagger f_i$}, so that the $\Uone$ spin-rotation symmetry manifest in \eqref{eq:HJ} corresponds to $\Uone$ number conservation, and $\hat{S}:\sop^z \to -\sop^z$ is implemented by $\hat{S}:f_{i} \to (-1)^i f_{i}^\dagger$, where $(-1)^i = \pm$ on A (B) sublattices.
 
The model \eqref{eq:HJ} has an extensive number of conserved quantities \cite{ufps20}, which correspond to the Wilson loop operators $W_\square = \prod_{\langle ij \rangle \in \square} u_{ij}$ of the $\Ztwo$ gauge field. Following Lieb's theorem, the ground state of Eq.~\eqref{eq:HJg} lies in the $\pi$-flux sector with $W_\square = -1 \forall \, \square$.
Note that there is a finite energy cost $\Delta  \sim J$ for flipping $W_\square = -1 \to +1$, which we henceforth refer to as the ``flux gap''.

Choosing a fixed gauge, depicted in Fig.~\ref{fig:introfic}, leads to a two-sublattice free-fermion Hamiltonian $\tilde{H}_J = 2 J \sum_{\bvec{k}} g(\bvec{k}) f_{\bvec{k},A}^\dagger f_{\bvec{k},B} + \hc$, where $g(\bm{k}) = 1 - \eu^{\iu \bm{k} \cdot \bvec{n}_1} + \eu^{-\iu \bm{k} \cdot \bvec{n}_2} + \eu^{\iu \bm{k} \cdot (\bvec{n}_1 - \bvec{n}_2)}$ is a function of the translation vectors of the chosen gauge, {$\bvec{n}_{1/2} = (1,\mp 1)$}.
Diagonalizing the latter reveals that the $f$-fermion bands features gapless Dirac cones at momenta $\bvec{\Sigma}_\pm = (0,\pm \pi/2)$.

We emphasize that, in a given gauge, physical symmetry transformations now act \emph{projectively}: a transformation $U$ is a \emph{physical} symmetry if it leaves $\tilde{H}_J$, Eq.~\eqref{eq:HJg}, invariant up to a gauge transformation $G_U$, i.e. $G_U^{-1} U^{-1} \tilde{H}_J U G_U = \tilde{H}_J$.
The set of all $\{G_U U\}$ spans the \emph{projective symmetry group} (PSG) \cite{wen02}, which can be used to characterize distinct parton theories.
Under these projective lattice operations, the Hamiltonian $\tilde{H}_J$ features a one-site unit cell and all symmetries of the original Hamiltonian \eqref{eq:HJ}:

(i) in-plane $\Uone$ spin rotation symmetry, (ii) a $\Ztwo$ Ising ``spin-flip'' symmetry $\hat{S}:\mathsf{s}^z \to - \mathsf{s}^z$ (which can be unitarily implemented via a spin rotation $C_2^y$ along the $\hat{y}$  axis in spin space), and (iii) time-reversal symmetry $\mathcal{T}$, acting as $\sop^\alpha \to - \sop^\alpha$ and, due to anti-unitarity, $\oop^y \to -\oop^y$.
Further, (iv), the Hamiltonian~\eqref{eq:HJ} respects \emph{all} space group (p4m) symmetries of the square lattice \cite{zerf19}, if the corresponding lattice symmetry operations are accompanied by appropriate transformations of the orbital degrees of freedom.

The projectively implemented symmetry operations are not necessarily local in momentum space \cite{SuppMat}. Consider, e.g., the action of translations on the fermionic parton operators{, e.g., on the $A$-sublattice in unit cell with coordinates $i_1 \bvec{n}_1 + i_2 \bvec{n}_2$}:
\begin{equation} \label{eq:T_x_stag}
    G_{T_x} T_x: f_{i_1 \bvec{n}_1 + i_2 \bvec{n}_2,A} \to (-1)^{i_1 + i_2} f_{(i_1+1) \bvec{n}_1 + i_2 \bvec{n}_2,B}
\end{equation}
so that $G_{T_x} T_x$ relates fermions at the Dirac point $\bvec{\Sigma}_\pm$ onto $\bvec{\Sigma}_\mp$, because above implies $G_{T_x} T_x: f_{\bvec{k},A} \to - \eu^{\iu \bvec{k}\cdot\bvec{n}_1} {f_{\bvec{k}+(\pi,0),B}}$ (we omit writing $G_U$ henceforth: ``symmetry'' shall refer to its respective projective implementation).
Anti-unitary time-reversal symmetry acts projectively as $\mathcal{T}^{-1} f_{\bvec{k},s} \mathcal{T} = \sigma^z_{s,s'} f_{\bvec{k},s'}^\dagger$ where $s=A,B$ sublattices.

{\em Altermagnetism: projective symmetry analysis.---}
Equipped with the above exactly solvable model of a spin\mbox{(-orbital)} liquid, we now investigate to what extent such a system may exhibit \emph{altermagnetism}.
We adopt a symmetry-based definition of an altermagnet as a \emph{phase of compensated local moments, where the magnetic sublattices are neither related by a translation nor by an inversion symmetry, but instead some other point group operation, e.g. rotations} \cite{mcclarty24}.

Henceforth, we systematically search for perturbations that induce altermagnetism.
We allow the symmetries (i)--(iv) to be broken but demand that the system retains (a) a combination of plaquette-centered fourfold lattice rotation operation $C_4^{(p)}$ and the Ising symmetry $\hat{S}$, as well as (b) a lattice reflection symmetry $R_x: (x,y) \to (-x,y)$ along the $\hat{y}$$\hat{z}$-plane through a site.

Assuming that perturbations are small compared to the flux gap $\Delta$, we focus on the Dirac fermions as the only low-energy degrees of freedom and write down a general low-energy Hamiltonian up to quadratic order in $\bvec{q}=(q_x,q_y)^\top$ (relative to the Dirac points at $\bm{\Sigma}_\pm$) as
\begin{equation}
    H^\mathrm{eff}(\bvec{q}) = \sum_{\mu,\nu=0,\dots,3}\left(a_{\mu \nu} + b_{\mu \nu}^i q_i + c_{\mu \nu }^{ij} q_i q_i \right)  \tau^\mu \sigma^\nu,
\end{equation}
where $\sigma^\mu = (\mathds{1}, \sigma^\alpha)$ and $\tau^\mu$ act on sublattice and valley degrees of freedom, respectively.
The coefficients $a_{\mu \nu}, b^i_{\mu \nu}$ and $c_{\mu \nu}^{ij}$ are constrained by demanding invariance under both (a) projectively implemented $C_4^{(p)} \hat{S}$ and (b) $R_x$.
The resulting invariant terms are listed in Tab.~\ref{tab:eff-symm-allowed}, along with their character under the individual $C_4^{(p)}$ rotations and spin-flip $\hat{S}$, time reversal $\mathcal{T}$ and translations $T_x,T_y$.
\begin{table}[htb]
\begin{tabular}{l||c|c|c|c }
& $C_4^{(p)}$ & $\hat{S}$ & $\mathcal{T}$ & $T_a$ \\
\hline \hline
Allowed by by $C_4^{(p)} \hat{S}$ and $R_x$ & & & &  \\
    \hspace{1.25em} \quad$(-q_x \tau^z +q_y^2) \sigma^x - (q_y + q_x q_y \tau^z)\sigma^y $ & $+1$ & $+1$ & $+1$ & $+1$ \\
    \hspace{1.25em} \quad $q_y \sigma^x + q_y^2 \sigma^y + q_x q_y \sigma^x \tau^z - q_x \sigma^y \tau^z$ & $-1$ & $-1$ & $+1$ & $+1$ \\
    A$_1^+$ \quad $\mathds{1}$ & $+1$ & $-1$ & $-1$ & $+1$ \\
    A$_1^-$ \quad $\sigma^z$, $(q_x^2 + q_y^2) \sigma^z$ & $-1$ & $-1$ & $-1$ & $-1$ \\
    B$_1$ \quad$(q_x^2 - q_y^2)\mathds{1}$ & $-1$ & $-1$ & $-1$ & $+1$ \\
    B$_2$ \quad$q_x q_y \tau^z$ & $+1$ & $+1$ & $-1$ & $-1$ \\
\hline
\hline
Order parameters & & & & \\
    A$_1^+$\quad Magnetization $m$, Zeeman field $h$ & $+1$ & $-1$ & $-1$ & $+1$ \\
    A$_1^-$ \quad Néel antiferromagnetism $n$ & $-1$ & $-1$ & $-1$ & $-1$\\
    B$_1$ mod A$_1^-$ \ Staggered lattice distortion $\kappa$ & $+1$ & $+1$ & $+1$ & $-1$ \\
    B$_2$ \quad Orbital antiferromagnetism $\lambda$ & $+1$ & $+1$ & $-1$ & $-1$
\end{tabular}
\caption{\label{tab:eff-symm-allowed}Terms in the low-energy Hamiltonian at the $\bvec{\Sigma}_\pm$-valleys allowed by projectively implemented $C_4^{(\mathrm{p})} \hat{S}$ and $R_x$ symmetries, and corresponding order parameters.
}
\end{table}

The first entry is fully symmetric. An explicit calculation shows that this is the gradient expansion of $\tilde{H}_J$, yielding the linear dispersion $\varepsilon^\pm_\tau =  \pm \tilde{J} |\bvec{q}|$ near $\bvec{\Sigma}_\pm$ where $\tau = \pm 1$ is the valley index{, and we write $\tilde{J} = 2 J$.}
The second entry in Tab.~\ref{tab:eff-symm-allowed} is found to simply renormalize the dispersion and will be neglected henceforth.

Next, consider the terms labeled ``A$_1^+$'' and ``A$_1^-$'', which transform exactly as the uniform magnetization $m$ and the Néel order parameter $n$, respectively.
While the former acts a chemical potential for the fermionic degrees of freedom, the latter enters the effective Hamiltonian as $H^\mathrm{eff}(\bvec{q}) \sim  n \sigma^z$, corresponding to a trivial mass term for the Dirac fermions $\bvec{\Sigma}_\pm$ and opening up a gap as $\varepsilon_\tau^\pm (\bvec{q}) \approx \pm \left(|n| + \frac{\tilde{J}^2}{2 |n|} |\bvec{q}|^2 \right)$.
Note that $(q_x^2 + q_y^2)\sigma^z$  may be thought of as its ``extended $s$-wave'' harmonic; it will therefore be ignored.

We now discuss the terms ``B$_1$'' and ``B$_2$'':
First, note that the ``B$_1$'' term, given by $(q_x^2 - q_y^2)$, has the same character under $C_4^{(p)}$, spin-flip $\hat{S}$ and time-reversal symmetry $\mathcal{T}$ as the product $n\kappa$ of the Néel-order parameter $n$ with a staggered lattice distortion field $\kappa$. Thus, there exists a symmetry-allowed coupling $H^\mathrm{eff}(\bvec{q}) \sim \kappa n (q_x^2 - q_y^2)$.

In contrast, the ``B$_2$''-term $q_x q_y \tau^z$ is odd under time-reversal symmetry and primitive translation, but even under $\hat{S}$. It, therefore, couples to an order parameter $\lambda$ for \emph{orbital antiferromagnetism}.

We deduce that if either $\kappa \neq 0$ or $\lambda \neq 0$ (or both $\kappa, \lambda \neq 0$) the system, described by the low-energy Hamiltonian $H^\mathrm{eff} = H^\mathrm{eff}_J + H^\mathrm{eff}_n + H^\mathrm{eff}_\kappa + H^\mathrm{eff}_\lambda$, is an \emph{altermagnet} for any finite Néel order parameter $n \neq 0$ in the sense that the ordered state preserves the combination $C_4^{(p)} \hat{S}$ of fourfold-lattice rotations $C_4^{(p)}$ with Ising spin-flip $\hat{S}$ while the combination $T_a \hat{S}$ of primitive translations $T_a$ with $\hat{S}$ is broken.

The Hamiltonian $H^\mathrm{eff}$ should be understood as an effective low-energy theory that contains all symmetry-allowed terms generated under a renormalization-group flow.
We distinguish between three separate fractionalized altermagnetic (``AM$^\ast$'') states:

\emph{First}, if both A$_1^-$ and B$_1$ are present (because of a finite Néel order parameter), but $\lambda = 0$, the system retains an additional $\mathcal{T} \hat{S}$-symmetry. We call this state ``AM$_1^\ast$''.
\emph{Second}, if $n \neq 0$, but the B$_1$-term vanishes ($\kappa =0$), altermagnetism is induced by orbital antiferromagnetism $\lambda \neq 0$ (inducing the B$_2$-term).
This state, dubbed ``AM$_2^\ast$'', breaks $\mathcal{T} \hat{S}$, but possesses an additional $\mathcal{T} T_a$-symmetry, as visible from Table~\ref{tab:eff-symm-allowed}.
\emph{Third}, ``AM$_3^\ast$'' refers to a state where A$_1^-$, B$_1$ and B$_2$ are finite, which does not possess any additional symmetries (apart from the imposed $C_4^{(p)} \hat{S}$ and $R_x$).

{\em Altermagnetism and parton band degeneracies.---} %

We now elucidate how the fermionic parton band structure exhibits a momentum-dependent breaking of time-reversal symmetry as a result of altermagnetism.
The parton band structure is straightforwardly obtained by diagonalizing $H^\mathrm{eff}(\bvec{q})$ and expanding to quadratic order,
\begin{equation}\label{eq:low-energy-disp}
    \varepsilon_{\tau}^\pm (\bvec{q}) \approx -\tau \lambda q_x q_y - \kappa n (q_x^2 - q_y^2) \pm \left( |n| + \frac{\tilde{J}^2}{2|n|} |\bvec{q}|^2\right).
\end{equation}
These low-energy bands are shown in Fig.~\ref{fig:introfic}(b) for AM$_1^\ast$ and Fig.~\ref{fig:introfic}(c) for all three distinct AM$^\ast$ phases.

The hallmark signature of conventional altermagnetism is a momentum-dependent \emph{spin splitting} of electronic or magnon bands.
However, in the model at hand, the spin degree of freedom is now encoded in the particle-hole character of the fermionic parton bands:

First, consider Néel order with $T_x \hat{S}$-symmetry. Taken together with inversion, this ensures particle-hole symmetric parton bands $\varepsilon^{\pm}_\tau(\bvec{q}) \overset{(T_x \hat{S})\text{-sym}}{=} - \varepsilon^{\mp}_{\tau}(-\bvec{q}) \overset{\mathcal{I}\text{-sym}}= -\varepsilon^\mp_\tau(\bvec{q})$, as visible in Fig.~\ref{fig:introfic}(b).
This is in analogy to fully spin-degenerate bands in conventional $T_x \hat{S}$-symmetric antiferromagnets.

In contrast, in fractionalized altermagnets, the parton band structure is no longer particle-hole symmetric. However, the system retains the $C_4^{(p)}\hat{S}$-symmetry: as illustrated in Fig.~\ref{fig:introfic}(b), a spin-flip operation (i.e. exchanging particle- and hole bands) yields a $C_4^{(p)}$-rotated version of the initial bands. 
Formally,
\begin{equation} \label{eq:c4-constraint}
    \varepsilon^{\pm}_\tau(\bvec{q}) \overset{(C_4^{(p)} \hat{S})\text{-sym}}{=} - \varepsilon^{\mp}_{\tau}(-q_y,q_x),
\end{equation}
ensuring a ``$d$-wave'' particle-hole splitting.
Importantly, we see that at the Dirac points, $\bvec{q} = 0$, the spectrum retains particle-hole--symmetry, analogous to the absence of a net ($s$-wave) spin splitting in non-fractionalized altermagnets.

While the above argument holds for all three AM$^\ast$ phases, the additional $\mathcal{T} \hat{S}$, and $\mathcal{T} T_x$, symmetries in AM$_1^\ast$, and AM$_2^\ast$ respectively, further constrain parton spectra.
They enforce particle-hole degeneracy along \emph{nodal lines} (see also \cite{libor22a,antonenko25,fernandes24}) for $q_x = \pm q_y$ in AM$_1^\ast$ [$(q_x,0)$ and $(0,q_y)$ in AM$_2^\ast$], as illustrated in Fig.~\ref{fig:introfic}(c).

{\em Microscopic origin of altermagnetic couplings.---}%
The above discussion was based on general symmetry principles that govern the low-energy physics of our model.
Remarkably, we are able to construct microscopic perturbations to Eq.~\eqref{eq:HJ} that do not spoil the model's exact solvability, {preserve the $\pi$-flux ground state sector up to intermediate coupling strengths,} and at low energies reduce to the effective Hamiltonian $H^\mathrm{eff}(\bvec{q})$. {These \emph{exactly solvable models} for fractionalized altermagnets are given explicitly in Eqs.~\eqref{eq:h-kappa-LGT} and \eqref{eq:h-lambda-LGT} in the End Matter.}

{\em Physically observable signatures.---}%
We stress that the spinless $f$-fermions carry a $\Ztwo$ gauge charge, and therefore, their single-particle spectra are not observable (we emphasize, however, that symmetry-based arguments, also pertaining to degeneracies, are gauge-invariant due to our use of projective symmetries).
This is in contrast to conventional altermagnets, where spin-split electron/magnon bands could be accessed using spin-polarized probes \cite{mcclarty25}.
Instead, consider the longitudinal dynamical spin-structure factor $\mathcal{S}^{zz}(\bvec{Q},\omega) = \frac{1}{V} \sum_{i,j} \int \du t \, \eu^{\iu \omega t} \eu^{-\iu \bvec{Q} \cdot (\bvec{r}_i - \bvec{r}_j)} \left\langle \sop^z_i(t)\sop^z_j(0) \right\rangle$.
Since $\sop^z$ does not induce flux excitations of the emergent gauge field, $\mathcal{S}^{zz}(\bvec{Q},\omega)$ can be obtained in the $\pi$-flux ground state sector from the density-density susceptibility $\chi(\iu \omega, \bvec{Q})$ of the fermionic partons $f$ as $\mathcal{S}^{zz}(\bvec{Q},\omega) = - 2 (1+n_\mathrm{B}(\omega)) \Im[\chi(\iu \omega \to \omega+ \iu 0^+,\bvec{Q})]$.
For frequencies slightly above the gap $\omega \gtrsim 2 |n|$, the dominant contributions are given by \emph{intravalley particle-hole excitations} at $\bvec{Q} \approx 0$ as well \emph{intervalley} excitations at $\bvec{Q} \approx (\pm \pi,0), (0,\pm \pi)$. 

\begin{figure}[tb]
    \centering
    \includegraphics[width=0.95\linewidth]{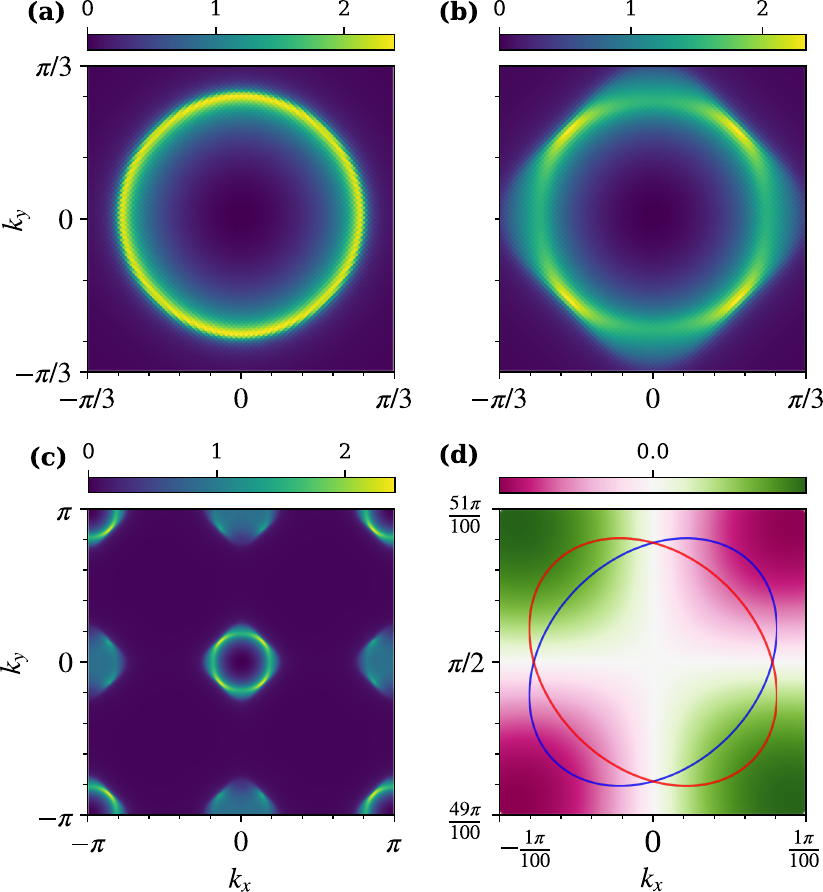}
    \caption{Constant-frequency cut ($\omega = 2.5 J$) of the intravalley contribution to the structure factor $\langle \sop^z \sop^z \rangle(\omega,\bvec{k})$ for (a) $\kappa= \lambda =0, n=0.3J$ and (b) $\lambda = 0,  n = 0.3J, \kappa n=0.6J$ {for which we also show the structure factor in the entire Brillouin zone in panel (c)}. (d) Berry curvature quadrupole at the $\bvec{\Sigma}_+$-valley. Blue (red) ellipses denote the parton Fermi surfaces for a positive (negative) applied magnetic field $h$ which are distorted by $\lambda$. }
    \label{fig:observables}
\end{figure}

In the $T_x \hat{S}$-symmetric antiferromagnet ($n\neq 0, \lambda=\kappa =0$), the intravalley contributions to $\mathcal{S}^{zz}$ exhibits a ring-like intensity distribution shown in Fig.~\ref{fig:observables}(a) due to circularly symmetric $\varepsilon(\bvec{q}) = \varepsilon(|\bvec{q}|)$ at low energies [see also Fig.~\ref{fig:introfic}(b)].
In the altermagnetic states with $\lambda \neq 0$ and/or $\kappa \neq 0$, the circular symmetry of $\mathcal{S}^{zz}$ is lifted, and instead ``hot spots'' emerge, shown in Fig. \ref{fig:observables}(b). {[The full structure factor, with intervalley contributions, is shown for the AM$^\ast_2$ state in Fig.~\ref{fig:observables}(c).]}
In all three (AM$_1^\ast$, AM$_2^\ast$ and AM$_3^\ast$) states $\mathcal{S}^{zz}(\omega,\bvec{Q})$ retains a fourfold rotational symmetry.
This is consistent with $\hat{S} C_4^{(p)}$-symmetry, since $\left\langle \sop^z_i(t)\sop^z_i(0) \right\rangle$ is invariant under $\hat{S} : \sop^z \to - \sop^z$.
{We have further computed the sublattice-resolved structure factor, which resolves the breaking of $\hat{S} T_a$-symmetry in the AM$^\ast$ phases \cite{SuppMat}.}

Having established $C_4^{(\mathrm{p})} \hat{S}$ as a fourfold rotational symmetry, consider applying a magnetic field $h$ along $\hat{z}$ which breaks $C_4^{(\mathrm{p})} \hat{S}$-symmetry ($h$ acts as a chemical potential on the $f$-partons).
Remarkably, this \emph{out-of-plane} field results in anisotropic \emph{in-plane} transport, for example in the thermal conductivity, $\kappa^{xx} \neq \kappa^{yy}$ {(for sufficiently low temperatures, (gapped) flux excitations of the $\Ztwo$ gauge fields are exponentially suppressed and their contribution to transport can be neglected)}.
This property is similar to conventional metallic AMs, where electrical and heat conductivities become anisotropic in an external magnetic field.
While these arguments apply for AM$_1^\ast$ and AM$_3^\ast$, the combination of $(C_4^{(\mathrm{p})} \hat{S}) (T_x \mathcal{T})$ in AM$_2^\ast$ remains a symmetry for finite $h \neq 0$, ensuring \emph{isotropic} $\kappa^{xx} = \kappa^{yy}$ also under applied fields.

Next, due to the global $\Uone$ spin-rotation symmetry (number conservation of $f$-fermions) one may study \emph{spin transport} {in response to gradients of the Zeeman magnetic field $\nabla^a h$ which act as an effective electric field on the $f$-fermions (see also End Matter). Both the field (and its gradients) do not spoil the conservation of the gauge field: hence, no $\Ztwo$ flux excitations are induced, and spin transport is mediated only by the fermionic partons.}

{As above, a finite \emph{uniform} background field $h =\mathrm{const.}$ breaks the $C_4^{(\mathrm{p})} \hat{S}$ and thus leads to anisotropic longitudinal spin conductivities, $\sigma^{xx} \neq \sigma^{yy}$ in the AM$_1^\ast$ and AM$_3^\ast$ phase. Note that, in}
the spirit of ``ideal altermagnetism'' (absence of spin-orbit coupling), $h$ is treated as a \emph{Zeeman}-field (scalar under spatial symmetries).
Then, $R_x:(x,y) \to (-x,y)$ remains a symmetry of the system and imposes vanishing spin Hall conductivities even in the presence of finite uniform fields, $\sigma^{xy}_\mathrm{Hall}(h) \equiv 0$.

{We now study non-linear spin transport, which may resolve} altermagnetic band splittings \emph{and} opposing Berry curvature in the two valleys, {and thus provides finite signatures for all three AM$^\ast$ states.} 
Nonlinear spin conductivities relate the current response $\bvec{J}^d$ to $\nabla^a h \nabla^b h \nabla^c h$ as \cite{sodemann15,Fang24}
\begin{align}
    \sigma^{abc;d} = -\frac{{e^*}^4}{\hbar}\sum_{i=0}^{3} \left ({\frac{-\tau}{\hbar}} \right )^i {\sigma}^{abc;d}_i,
\end{align}
where $\tau$ is the scattering time{, $e^*$ is an effective charge (cf. End Matter)} and the ${\sigma}_{i}$ contributions are given in the SM \cite{SuppMat}.
Symmetry arguments and our explicit calculations \cite{SuppMat} show that
\begin{equation} \label{eq:nonlinear-results} 
    \sigma^{xyx,y}_2(h) \sim \lambda n h \quad \text{and} \quad \sigma^{xxx,x}_1(h)-\sigma_1^{yyy,y}(h) \sim \kappa n h,
\end{equation}
where the first term arises due to a Berry curvature quadrupole $\sim \partial_x \partial_y \Omega_{xy}$, and the second term due to a quantum metric quadrupole \cite{zhang23,Fang24}.
Here, the quantum metric and Berry curvature are independent of $\lambda$ and/or $\kappa$: instead, the terms arise from the altermagnetic distortion of $f$-fermion bands.
We conclude that the altermagnetism due to $\lambda \neq 0$ (AM$_1^\ast$ and AM$_3^\ast$) {leads to} anisotropic longitudinal transport, while  $\kappa \neq 0$ (in AM$_2^\ast$ and  AM$_3^\ast$) results in a nonlinear spin Hall conductivity, {thus allowing for an observable distinction between AM$_1^\ast$, AM$_2^\ast$ and AM$_3^\ast$ states}.

{\em Conclusion.---}%
We have constructed exactly solvable spin-orbital models which realize \emph{fractionalized altermagnetism} and shown that the momentum-dependent spin splitting is, remarkably, encoded in a broken particle-hole degeneracy of fractionalized fermion degrees of freedom, and discussed physical observables.
Further studies could investigate deconfined systems with distinct fractionalization patterns and explore mechanisms for realizing spin-orbital liquids and their altermagnetic deformations in transition-metal compounds \cite{churchill25} and quantum simulators \cite{verresen22}.

{\em Note added.---}
During the preparation of this work, Ref.~\onlinecite{vijayvargia25} appeared, which also studies altermagnetism in an exactly solvable spin-orbital model, on bilayers of the square-octagon lattice.
%

\acknowledgments

{\em Acknowledgments.---}We would like to thank O.~Erten and collaborators for sharing their preprint \cite{vijayvargia25} ahead of arXiv submission. We gratefully acknowledge discussions with A.~Joy, E.~J.~König and M. Scheurer.
This work is funded by the Deutsche Forschungsgemeinschaft (DFG, German Research Foundation) through SFB 1238, project id 277146847 (UFPS and AR), and the Emmy Noether Program, project id 544397233, SE 3196/2-1, (UFPS). J.K. and A.N. acknowledge support from the Imperial-TUM flagship partnership, the Deutsche Forschungsgemeinschaft (DFG, German Research Foundation) under Germany's Excellence Strategy--EXC--2111--390814868, DFG grants No. KN1254/1-2, KN1254/2-1, and TRR 360 - 492547816 and from the International Centre for Theoretical Sciences (ICTS) for the program ``Frustrated Metals and Insulators'' (code: ICTS/frumi2022/9), as well as the Munich Quantum Valley, which is supported by the Bavarian state government with funds from the Hightech Agenda Bayern Plus.

\appendix

{\section{End Matter}

{\em Microscopic spin-orbital realizations of altermagnetic deformations.---} In this section, we provide exactly solvable perturbations to Eq.~\eqref{eq:HJ} which in the $\pi$-flux sector give rise to the low-energy parton dispersion \eqref{eq:low-energy-disp}:
First, the Néel order parameter corresponds to a staggered field with $H_n = - n \sum_{i} (-1)^i \sop^z_i$, which using {$\sop^z_i = -2 f_i^\dagger f_i + 1$} can be easily seen to give rise to the A$_1^-$ term.
Second, Hamiltonians that give rise to $H^\mathrm{eff}_\kappa$, and $H^\mathrm{eff}_\lambda$ respectively, are given by 
\begin{subequations}
    \begin{align}
        H_\kappa &=  n \kappa \sum_{s = \pm} \sum_i s u_{i,i+\bvec{\delta}_s} u_{i+\bvec{\delta}_s,i+2\bvec{\delta}_s} (f_i^\dagger f_{i+2\bvec{\delta}_s} + \hc) \label{eq:h-kappa-LGT} \\
        H_\lambda &=  \lambda \sum_{\substack{\circlearrowright{}_{i \in A}\\ \circlearrowleft_{i \in B}} \langle ijk \rangle} \iota_{ik}  u_{kj} u_{ji} (\iu f_k^\dagger f_{i} + \hc) \label{eq:h-lambda-LGT},
\end{align}
\end{subequations}
where $\bvec{\delta}_\pm = \frac{\bvec{n}_1\pm \bvec{n}_2}{2}$ and $\circlearrowright \langle ijk \rangle $ refers to the clockwise summation over three sites within the same plaquette and {$\iota_{ik}$ is $\pm 1$ if $k-i = \substack{\bvec{n_1}\\{\bvec{n_2}}}$ otherwise zero}.
By writing $u_{ij}$ and $f$ in terms of Majorana degrees of freedom, one may then find corresponding spin-orbital Hamiltonians which read
\begin{subequations}
\begin{align}
    H_\kappa =  \frac{n \kappa }{2} \Big [&\sum_{\rightarrow \langle ijk \rangle ^{\gamma \gamma^\prime} }(  \epsilon^{\alpha z \beta} \sop^\alpha_{ i}  \sop^\beta_k) (\epsilon^{\gamma z \gamma^\prime}  \oop^\gamma_{ i} \oop^{z}_{ j}\oop^{\gamma^{\prime }}_{ k}) \nonumber\\  +&\sum_{\substack{\uparrow{}_{i \in A}\\ \downarrow_{i \in B}} \langle ijk \rangle} \bvec{\mathsf{s}}_i \cdot \bvec{\mathsf{s}}_k (\sop^z_{j} \oop^z_{i}   \oop^z_{j}) \Big ] \label{eq:h-kappa-so}
\end{align}
    \begin{align}
        H_\lambda &=  \frac{\lambda}{2} \sum_{\substack{\circlearrowright{}_{i \in A}\\ \circlearrowleft_{i \in B}} \langle ijk \rangle^{\gamma \gamma^\prime}}     \iota_{ik}   (\bvec{\sop}_{ i} \cdot \bvec{\sop}_{ k}) (\epsilon^{\gamma \eta \gamma^\prime} \oop^\gamma_i \oop^\eta_j \oop^{\gamma^\prime}_k), \label{eq:h-lambda-so}
\end{align}
\end{subequations}
where $\rightarrow \langle ijk \rangle ^{\gamma \gamma^\prime}$ refers to a next-nearest neighbour pair $ik$ with $\bvec{r}_k-\bvec{r}_i = \bvec{n}_1+\bvec{n}_2$ via the common nearest neighbour $j$ with bond variable $\gamma (\gamma^\prime)$ on the $ij$ ($jk$) bond, and analogously for the other directions.
Note that these terms may not be present in a ``bare'' Hamiltonian, but can be generated by parton-parton 
interactions under a renormalization-group flow (see, for example, Ref.~\onlinecite{ufps20}, where $H_n$ is shown to be generated by sufficiently strong nearest-neighbor Ising spin-spin interactions).
For more details on the parton construction and the derivation of Eqs.~\eqref{eq:h-kappa-so} and \eqref{eq:h-lambda-so} from Eqs.~\eqref{eq:h-kappa-LGT} and \eqref{eq:h-lambda-LGT}, we refer the reader to the Supplemental Material \cite{SuppMat}.

While the $\pi$-flux sector is stable against weak $\kappa n \ll J$ and $\lambda \ll J$, sufficiently strong deformations may lead to transitions to other flux sectors. We have therefore determined the energy of $H = H_J + H_n + H_\kappa + H_\lambda$ in various flux sectors as a function of $\kappa n$ and $\lambda$. Our results \cite{SuppMat} show that the ground state of $H$ lies in the $\pi$-flux sector until these deformations are approximately of order $J$, i.e.~$\mathrm{min}(n \kappa, \lambda) \sim J$, justifying our low-energy approach/analysis of the AM$^\ast$ states in a wide parameter regime.

{\em Zeeman field and its gradient as $f$-fermion chemical potential and electric field.---}%
From $\sop^z = 1 -2 f^\dagger f$, it follows that the Zeeman field $h$ in $H_h = - h \sum_i \sop^z_i$ couples linearly to parton number operator $n = f^\dagger f$. A uniform Zeeman field $h_0$ thus corresponds to an effective chemical potential $\mu = -2 h_0$ for the fermionic $f$-partons.
Now consider a spatially varying Zeeman field $h_i = h(\bvec{r}_i)$, where $\bvec{r}_i$ denote the coordinates of the square lattice site $i$. If the field varies slowly, we may gradient-expand to first order, writing $h(\bvec{r}_i) = h_0 + \bvec{r}_i \cdot (\bvec{\Delta} h)_0 + \dots$, where $(\bvec{\Delta}h)_j = \left(h(\bvec{r}_j + \hat{x}) - h(\bvec{r}_j), h(\bvec{r}_j + \hat{y}) - h(\bvec{r}_i) \right)^\top$ is the lattice gradient evaluated at site $j$. Thus, $H_h = - \sum_i h_i \sop^z_i$ can be written as
\begin{align}
    H_h &= - \sum_i h_0 (-2) n_i - \sum_i (-2) (\bvec{\Delta} h)_0 \cdot \bvec{r}_i n_i + \dots \nonumber\\
    &= - \mu n - \bvec{P} \cdot \bvec{E} + \dots,
\end{align}
where $n = \sum_i n_i$ is the total $f$-fermion number, and we have defined the effective polarization $\bvec{P} \equiv e^\star \sum_i  \bvec{r}_i n_i$ which couples to an effective electric field $\bvec{E} \equiv (\bvec{\Delta} h)$ with coupling constant $e^\ast = -2$.

We conclude that gradients of the Zeeman field $h$ act as an electric field for the $f$-partons, and moreover \emph{do not} spoil the conservation of the $\Ztwo$ gauge field (since both the local fermion number $n_i$ and dipole operator $\bvec{r}_i n_i$ commute with $u_{ij}$ and thus also with $W_\square$).
Consequently, one may use standard methods (semiclassical Boltzmann equation, expanded in orders of $\bvec{E}$, in the relaxation-time approximation \cite{sodemann15,Fang24}, or equivalently (non-)linear response which involves an evaluation of current-current correlators (bubble diagrams) with phenomenological lifetime $\tau$ to capture disorder effects) to obtain the current $\bvec{J} = \partial_t \bvec{P}$ response of the fermionic partons to an electric field (i.e. gradient in the Zeeman field) in the $\pi$-flux background configuration, which then directly determines the linear and non-linear spin conductivities.
Note that we evaluate the corresponding conductivities as response functions to applied gradients at a given uniform background fields (chemical potential), and we have verified that the AM$^\ast$ states remain stable against finite uniform $h$ (i.e. no transitions to other flux sectors occur) \cite{SuppMat}.
We refer the reader to the Supplemental Material \cite{SuppMat} for explicit expressions of the respective (non-)linear conductivities, their relation to band-geometric quantities, and a numerical evaluation of these.

}

\clearpage
\ifarXiv
    \foreach \x in {1,...,\numbersupplementpages}
    {
        \includepdf[pages={\x,{}}]{\supplementfilename}
    }
\fi 
\end{document}